\documentclass[%
 aip,
 amsmath,amssymb,
 reprint,%
]{revtex4-1}

\usepackage{graphicx}
\usepackage{dcolumn}
\usepackage{bm}

\usepackage[utf8]{inputenc}
\usepackage[T1]{fontenc}
\usepackage{mathptmx}

\def\veck{\mathbf k}
\def\vecq{\mathbf q}
\def\vecQ{\mathbf Q}

\def\be{\begin{equation}}
\def\ee{\end{equation}}

\begin{document}

\preprint{AIP/123-QED}

\title[1D-Hubbard]{Antiferromagnetic fluctuations in the one-dimensional Hubbard model}

\author{V\'aclav  Jani\v{s} }
\email{janis@fzu.cz}
 \author{Anton\'\i n Kl\'\i\v c} \author{Jiawei Yan} 

\affiliation{Institute of Physics, The Czech Academy of Sciences, Na Slovance 2, CZ-18221 Praha  8,  Czech Republic}

\date{\today}

\begin{abstract}
We study the low-temperature critical behavior of the one-dimensional Hubbard model near half filling caused by enhanced antiferromagnetic fluctuations. We use a mean-field-type approximation with a two-particle self-consistency renormalizing the bare interaction. It allows us to control a transition from high to low temperatures as well as from weak to strong-coupling. We show that there is a crossover temperature $T_{0}= t\exp\{-1/U\rho(0)\}$ for arbitrary interaction $U>0$ and the bare density of states at the Fermi energy $\rho(0)>0$. The solution at lower temperatures goes over to strong coupling and approaches a quantum critical point with the diverging staggered  susceptibility  and a gap in the excitation spectrum at zero temperature.    
\end{abstract}

\maketitle 
\section{Introduction}

The exactly solvable quantum many-body models set limiting situations for testing non-perturbative approximate solutions in the strong-coupling regime. There are three fundamental models with an exact low-temperature solution, the Kondo model,\cite{Andrei:1983aa}  the single-impurity Anderson model (SIAM),\cite{Tsvelick:1983aa} and the one-dimensional Hubbard model.\cite{Essler:2005aa}  The solutions of these models are baed on the quantum-mechanical Bethe Ansatz.\cite{Bethe:1931aa}  This algebraic method gives the rigorous analytic expressions for the ground-state properties. The behavior of the excited states at non-zero temperatures can then be determined from the Bethe Ansatz only approximately.\cite{Takahashi:1972aa} In particular, the dynamics of the elementary excitations can be obtained only with additional approximations. 

The most direct and versatile method for studying the dynamics of the many-body quantum systems are the Green functions and Feynman diagrams. Renormalizations must be introduced in the diagrammatic representations of physical processes so that to allow for the transition from the weak-coupling to the strong-coupling regime. The standard perturbation expansions use a small parameter but their renormalizations pick up only selected classes of contributions that can no longer be controlled by the expansion parameter. They may be, however, justified in specific limits of the studied models. The best way to prove reliability of specific self-consistent approximations is to compare them with the exact solution, when available.            

The SIAM in the strong-coupling Kondo regime is often used to check the reliability of approximations on the local quantum dynamics which led to the introduction of the dynamical mean-field theory (DMFT).\cite{Georges:1996aa} The quantum local models contain the full quantum dynamics but miss the effects of spatial fluctuations. The latter are important for the formation of the long-range order in low-dimensional models where one has to obey the Mermin-Wagner theorem prohibiting continuous transitions to ordered states at non-zero temperatures.\cite{Mermin:1966aa} Since this theorem does not  restrict the long-range order at zero temperature, reliable approximations in low dimensions must be able to distinguish between non-zero and zero temperatures. Most of the approximate approaches to strong-electron correlations fail to do so and either allow or prohibit transitions to ordered phases uniformly for all temperatures.       

The problem with the description of the low-energy excitations in the one-dimensional Hubbard model is that one cannot start with the weak-coupling Fermi liquid. The spectrum of the excitations in the half-filled band displays a charge gap and the low-lying excitations do not have the fermionic character. Instead, a Luttinger liquid, where spin and charge excitations are separated, uses bosonic excitations, spinons and holons.\cite{Haldane:1981ac} The critical behavior of the spin and charge correlation functions at zero temperature were determined.\cite{Frahm:1990aa,Schulz:1990aa} Using the thermodynamic Bethe Ansatz the homogeneous thermodynamic properties of the one-dimensional Hubbard model were also numerically determined.\cite{Kawakami:1989aa,Deguchi:2000aa,Takahashi:1999aa,Takahashi:2002aa}  Although antiferromagnetic short-range order near half filling was discussed,\cite{Shiba:1972aa,Shiba:1972ab} the low-temperature asymptotics of the staggered susceptibility measuring the range of antiferromagnetic correlations has not been determined yet. In particular, the way the Fermi liquid breaks down in the low-temperature limit in the half-filled band without being antiferromagnetically ordered has not been addressed within the many-body perturbation theory.  

We formulate the approximation with fermionic elementary excitations for non-zero temperatures where the weak-coupling theory and/or high-temperature expansion can be applied.\cite{Shiba:1972aa,Takahashi:2002aa}  To reach zero temperature one has, however, to use  self-consistent non-perturbative approximations where the solution at half filling goes over to an insulator at arbitrary positive interaction. We developed a general mean-field-like theory with two-particle self-consistency that suppresses spurious transitions of the weak-coupling mean-field theory.\cite{Janis:2007aa,Janis:2017aa,Janis:2019aa} It proved its reliability in reproducing qualitatively correctly the zero-temperature Kondo limit\cite{Janis:2017ab,Janis:2019aa} as well as a crossover from high to low-temperature regimes of the SIAM.\cite{Janis:2020aa} Most recently we disclosed the existence of the Curie-Weiss law in the magnetic susceptibility of the SIAM on an interval of intermediate temperatures for sufficiently strong electron correlations.\cite{Janis:2020ab}  The general approach is also applicable to extended lattice systems.\cite{Janis:2019aa} The one-dimensional Hubbard model offers now a possibility to check reliability of this approximation for extended systems. The approximation primarily aims at two-particle correlation functions and is generally justified in the critical region of a response function. Here we use it to determine the low-temperature limit of the staggered susceptibility in the charge-symmetric state of the one-dimensional Hubbard model. The antiferromagnetic fluctuations or spin-flip processes are decisive for the dynamics of the elementary excitations and opening of the charge gap in the spectral function.

\section{Critical antiferromagnetic fluctuations}

The Hamiltonian of the Hubbard model on a linear chain with $N$ lattice sites with periodic boundary conditions is
\begin{equation}\label{eq:H-SIAM}
  H_{I} =  N\sum_{\sigma}\int_{-\pi/ l_{0}}^{\pi/ l_{0}}\frac{l_{0}dk}{2\pi }\epsilon( k)
  c^{\dagger}_{k\sigma} c^{\phantom{\dagger}}_{k\sigma} 
  +  U\sum_{\mathbf{i}} c^{\dagger}_{\uparrow\mathbf{i}} c_{\uparrow \mathbf{i}} c^{\dagger}_{\downarrow \mathbf{i}}  c_{\downarrow \mathbf{i}} \,, 
\end{equation}
with the dispersion relation  
\be
\epsilon(k) = - 2t \cos(l_{0}k) \,
\ee
and $l_{0}$ being the interatomic distance. 

The ground state of this model was solved exactly.\cite{Lieb:1968aa} The most important result for the half-filled band was the absence of the Mott transition at nonzero interactions U. The ground state is an insulator for any $U>0$. It means that the weak-coupling expansion is not applicable and the low-temperature state becomes increasingly strongly coupled with decreasing temperature. The gap in the  excitation spectrum is not, however, caused by a long-range antiferromagnetic order. There is hence a mechanism driving the system to the insulating state without becoming antiferromagnetically ordered.        

It is evident that the antiferromagnetic fluctuations with dynamical spin-flip processes will have the most important impact on the low-temperature behavior for half filling with suppressed charge fluctuations. The general response of the model on weak external magnetic excitations is the dynamical susceptibility. If we assume that the bare interaction $U$ is renormalized, screened to an effective local and static interaction $\Lambda$, the dynamical susceptibility generally is
\begin{equation}\label{eq:chi-dynamic}
\chi(\vecq,z)= - \frac{2 \phi(\vecq,z)}{1 + \Lambda \phi(\vecq,z) }\,,
\end{equation}
where $z$ is a complex frequency and $\phi$ is the electron-hole  bubble that can be represented in the Matsubara formalism as a convolution of two propagators
\begin{equation}
\phi(\vecq,i\nu_{m}) = \frac 1N\sum_{\veck}\frac 1\beta\sum_{\omega_{n}}G(\vecq + \veck, i\omega_{n + m}) G(\veck,i\omega_{n}) \,,
 \end{equation}
 where $\omega_{n}= (2n + 1) \pi T$ are fermionic and $\nu_{m}= 2m \pi T$ are bosonic Matsubara frequencies at temperature $T$ and $\beta = 1/T$. We set the Boltzmann constant $k_{B} = 1$.  
 
 The full one-electron propagator is
 \begin{equation}\label{eq:G-full}
\mathcal{G}(\veck, z) = \frac1{z + \bar{\mu} - \epsilon(\veck) - \Sigma^{Sp}(\veck,z)} \,
\end{equation}
with the effective chemical potential measuring the distance from half filling $\bar{\mu} = \mu + Un/2$ and the spectral self-energy $\Sigma^{Sp}(\veck,z)$. The approximation is then determined by the dependence of the effective interaction $\Lambda$ and the self-energy $\Sigma^{Sp}(\veck,z)$ on the bare interaction $U$. We generally split the approximation into the two-particle part determining the homogeneous vertex $\Lambda$ and the one-particle part determining the spectral self-energy $\Sigma^{Sp}(\veck,z)$.  We introduce a thermodynamic propagator $G(\veck,i\omega_{n})$  with only the static, Hartree self-energy that we use in the two-particle part determining the relation between the bare interaction and vertex $\Lambda$. This restriction is chosen to derive analytic expressions for the two-particle functions.  
 
The static susceptibility at for $z=0$  has a maximum at  a momentum $\vecq = \vecQ$ determining the type of the low-temperature ordering. It is the edge vector  $Q=\pi/l_{0}$  for the one-dimensional half-filled Hubbard model and $\vecQ=(\pi/l_{0},\pi/l_{0},\ldots)$ in higher dimensions. We identified the momenta with the wave vectors and used $\hbar =1$. The electron-hole bubble with the Hartree Green functions can be analytically represented after summing over the Matsubara frequencies 
\begin{equation}\label{eq:EH-bubble-explicit}
\phi(\vecq,\omega_{+}) =  \frac 1N\sum_{\veck}\frac {f(\epsilon(\veck) - \bar{\mu}) - f(\epsilon(\veck + \vecQ + \vecq) - \bar{\mu})}{\omega_{+} + \epsilon(\veck) - \epsilon(\veck + \vecQ + \vecq)} \,,
\end{equation}
where $f(x) = 1/(1 + e^{\beta x})$ is the Fermi function. We abbreviated the limit of complex frequencies to the real axis $\omega_{\pm} = \omega \pm i0^{+}$.

Antiferromagnetic fluctuations are dominant at low temperatures around the half-filled band $n=1$ with $\bar{\mu} = 0$. They are controlled by small transfer momenta of the two-particle bubble around the edge vector $\vecQ$. It is then sufficient to expand the electron-hole bubble in small momenta around $\vecQ$ to describe the critical antiferromagnetic fluctuations,
\be
\phi(\vecQ + \vecq, \omega_{+}) \doteq \phi(\vecQ,\omega_{+}) + Dl_{0}^{2}q^{2}\,.
\ee 
We expand the dispersion relation and use only the first two terms
\be
\epsilon(\veck + \vecQ + \vecq) \doteq \epsilon(\veck + \vecQ)  + l_{0}\vecq\cdot\nabla\epsilon(\veck + \vecQ)
\ee
to determine the spatial expansion parameter $D>0$. We further use the symmetries $\epsilon(\veck + \vecQ) = -\epsilon(\veck)$ and the same one  for the derivative $\nabla\epsilon(\veck + \vecQ) = -\nabla\epsilon(\veck)$ for the edge vector $\vecQ=(\pi/l_{0},\pi/l_{0},\ldots)$ to transform the momentum integrals into energy integrals with the local density of states, that in the one-dimensional model reads
\be
\rho(\epsilon) = \frac 1{2\pi} \frac {\theta\left(4t^{2} - \epsilon^{2}\right)}{\sqrt{4t^{2} - \epsilon^{2}}}  \,.
\ee
Here $\theta(x)$ is the Heaviside step function. The homogeneous part of the electron-hole bubble then is 
\begin{multline}\label{eq:phi-AF}
\phi(\vecQ,\omega_{+}) = P\int_{-\infty}^{\infty} d\epsilon \rho(\epsilon) \frac{f(\epsilon - \bar{\mu}) - f(-\epsilon - \bar{\mu})}{2 \epsilon + \omega_{+} }  
\\
-\  \frac{ i\pi }2\ \rho\left(-\frac{\omega}2\right)) \left[f\left(-\frac{\omega}2 - \bar{\mu}\right) - f\left(\frac{\omega}2 - \bar{\mu}\right) \right] \,.
\end{multline}
and 
\begin{multline}
D = \int_{0}^{2t} d\epsilon \rho(\epsilon) \frac{4t^{2} - \epsilon^{2}}{4\epsilon}\left[\frac{f(\epsilon - \bar{\mu}) - f(-\epsilon - \bar{\mu})}{2 \epsilon^{2}}   
\right. \\ \left.
-\  \frac{f^{'}(-\epsilon - \bar{\mu})}{\epsilon} -  f^{''}(-\epsilon - \bar{\mu})\right] \,,
\end{multline}
with the prime denoting the derivative.  

\section{Effective interaction}

The gist of our approach is to determine a screening of the bare interaction due to multiple scatterings of electron pairs. The renormalized interaction is derived from the two-channel parquet equations\cite{Bickers:1991ab,Rohringer:2018aa} reduced so that to stay in the critical region of the response function during the transition from weak to strong coupling.\cite{Janis:2019aa} The critical behavior manifests itself as a singularity in the Bethe-Salpeter equation with multiple electron-hole scatterings in the models with the repulsive interaction.\cite{Janis:1999aa} The bare interaction in this equation must be replaced by a renormalized one to avoid spurious transition to a magnetically ordered state in strong coupling. It is achieved by multiple scatterings of electron pairs. The  corresponding Bethe-Salpeter equation for the effective interaction $\Lambda$ in this scheme is\cite{Janis:2019aa} 
\begin{multline}\label{eq:Lambda-full}
\left[1 - \frac {\Lambda^{2}}N\sum_{\vecq}\frac 1\beta\sum_{\nu_{m}} \phi(-\vecq,-i\nu_{m})
\right. \\ \left. 
\times\frac{G(\veck + \vecq,i\omega_{n + m}) G(\veck'- \vecq,i\omega_{n'- m})}{1 + \Lambda\phi(-\vecq,-i\nu_{m})}  \right] \Lambda = U \,.
\end{multline}
This equation cannot hold for the static vertex $\Lambda$ point-wise for all fermionic momenta and frequencies, unless we turn the vertex dynamical. A dynamical vertex would make the resulting approximation extremely complicated  and would lead to the loss of the analytic control of the critical behavior that is the primary objective of our approximation.  If we cannot satisfy Eq.~\eqref{eq:Lambda-full} fully we resort to an approximate solution in a mean sense. There are several ways to do that. The critical behavior is not qualitatively changed by fluctuations in the fermionic variables, since they are irrelevant in the critical region of the pole in the response function. Different ways of treating the fermionic fluctuations in Eq.~\eqref{eq:Lambda-full} do not change the universal critical behavior but affect the non-universal quantities. Here we multiply both sides of Eq.~\eqref{eq:Lambda-full} by the product of the one-particle thermodynamic propagators $G(\veck,i\omega_{n})G(\veck',i\omega_{n'})$ and sum/integrate over the fermionic variables $\veck,\omega_{n}$ $\veck',\omega_{n'}$. Equation~\eqref{eq:Lambda-full} with $n$ being the charge density per site reduces after analytic continuation to real frequencies to      
\begin{equation}\label{eq:Lambda-phi2a}
\Lambda = \frac{Un^{2}}{n^{2} + 4\Lambda^{2} \mathcal{X}_{\vecQ} } \,,
\end{equation} 
where
\begin{multline}\label{eq:X-phi2}
\mathcal{X}_{\vecQ} = - S_{d}l_{0}^{d}  P\int_{-\infty}^{\infty}\frac{d \omega}{\pi} b(\omega) 
\\
\times \int_{0}^{L} \frac{q^{d - 1}dq}{(2\pi)^{d} }\Im\left[ \frac{\phi(\vecQ,\omega_{+})^{3}}{1  + \Lambda\left(\phi(\vecQ,\omega_{+})   + Dl_{0}^{2}q^{2}\right)}\right] > 0 \,
\end{multline}
is the term renormalizing the bare interaction due to multiple scattering of singlet electron pairs.  
We denoted $b(x) = 1/(1e^{\beta x} - 1)$ the Bose function, $S_{d}$ is the surface of the $d$-dimensional unit sphere and $L$ is an appropriate cutoff for large momenta needed for dimensions $d>3$.  Positivity of integral $\mathcal{X}_{\vecQ}$ leads to a screening of the bare interaction. 

Equation~\eqref{eq:Lambda-full}  self-consistently  determines the effective interaction $\Lambda$ that controls the critical behavior near the pole of the magnetic response function, a singularity in the integrand in Eq.~\eqref{eq:X-phi2}. The singularity emerges  for $a = 1 + \Lambda\phi(\vecQ,0) =0$. This critical point can be reached only if $\mathcal{X}_{\vecQ}<\infty$. It means that the singularity must be integrable.  The self-consistent equation for the effective interaction then suppresses spurious poles of the random-phase approximation with the bare interaction $U$ and correctly allows only for integrable singularities in the response functions.     

Equations~\eqref{eq:phi-AF}-\eqref{eq:X-phi2} determine our general approximation of the mean-field character with a two-particle self-consistency that can be applied in any spatial dimension in the critical region of the diverging susceptibility. The asymptotic integral over the transfer momentum $q$ in Eq.~\eqref{eq:X-phi2} can be performed and its explicit value in the critical region of diverging antiferromagnetic fluctuations in one dimension is 
\begin{multline}\label{eq:X-omega}
\mathcal{X} =  -\frac{1}{2\pi \sqrt{\Lambda D}} \int_{0}^{2t} d\omega
\\
 \frac{\phi(0)^{3}\coth\left(\beta \omega/2\right)\sin\left(\alpha(\omega)/2\right)}{\left[ \left(1 + \Lambda \Re\phi(\omega_{+})\right)^{2} + \Lambda^{2}\Im\phi(\omega_{+})^{2}\right]^{1/4}} \,,
\end{multline}  
where $\alpha(\omega) = \arctan\left(-\Lambda\Im\phi(\omega_{+})/(1 + \Lambda \Re\phi(\omega))\right)$ and we left out the dependence on the fixed edge vector $Q=\pi/l_{0}$ to simplify the notation.  Equation~\eqref{eq:Lambda-phi2a} for the effective interaction $\Lambda$ together with equation~\eqref{eq:X-omega} for integral $\mathcal{X}$ can be solved numerically in the whole phase space of the input parameters in the same way as it was done in the SIAM.\cite{Janis:2019aa}

\section{Low-temperature asymptotics of the one-dimensional Hubbard model}

\subsection{General case}

The most interesting situation in the Hubbard model is the low-temperature limit of the half-filled band with enhanced antiferromagnetic fluctuations. It is manifested by proximity of a singularity in the static staggered susceptibility 
\begin{align}\label{eq:Chi-AF-static}
\chi^{AF} &= -\frac{2\phi(0)}{a} 
\end{align}
with the bubble $\phi(\omega)$ from Eq.~\eqref{eq:phi-AF} and a dimensionless Kondo scale $a= 1 + \Lambda\phi(0)$ measuring the distance from the critical point. We can use  equation~\eqref{eq:Lambda-phi2a} as an equation for the Kondo scale  $a\in (0,1)$  with $\Lambda= (a - 1)/\phi(0)$. 

The dominant contribution to integral $\mathcal{X}$ in the critical region with $a\ll 1$ comes from small frequencies and we can replace the bubble by its low-frequency asymptotics 
$\phi(\omega_{+}) \doteq \phi(0) - i\pi A\omega$. We obtain from Eq.~\eqref{eq:phi-AF}
\begin{align}
A &= \frac{\beta}2 f(\bar{\mu}) f(-\bar{\mu})\rho(0)\,. 
\end{align}
Moreover, we can set $\omega=0$ in all regular functions of the integrand in Eq.~\eqref{eq:X-omega}. We must, however, introduce a frequency cutoff. We choose the hopping parameter t to suppress the irrelevant contribution from high frequencies. We thereby do not  affect the leading-order critical asymptotics. We replace the integrand in  Eq.~\eqref{eq:X-omega} by its low-frequency asymptotics, that is     
\begin{multline}\label{eq:X-full}
\mathcal{X} = -\frac{\phi(0)^{3}}{2\pi \sqrt{\Lambda D}} \int_{0}^{2t}d\omega \frac{\coth\left(\beta \omega/2\right)\sin\left(\alpha(\omega)/2\right)}{\left(a^{2} + \pi^{2}\Lambda^{2}A^{2}\omega^{2}\right)^{1/4}} \,,
\end{multline}  
 with $\alpha(\omega) = \arctan(\pi\Lambda A\omega/a)$.
 
\begin{figure}
\includegraphics[width=8cm]{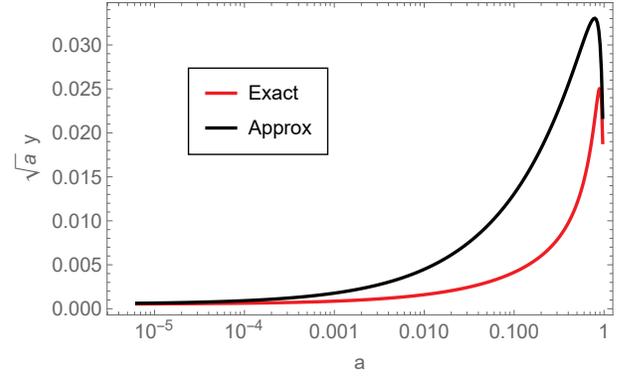}
\caption{Comparison of a dimensionless integral $\sqrt{a}y= \sqrt{a}\mathcal{X}/\phi(0)^{2}$ calculated with the asymptotic form, Eq.~\eqref{eq:X-asymptotic}, black line, and the full expression, Eq.~\eqref{eq:X-full}, red line, as a function of the Kondo parameter $a$ on the logarithmic scale for $U=5t$ and $T=0.01t$.   \label{fig:y-int-comp} }
\end{figure}

We can analytically assess the integral in the leading order of the vanishing scale $a\to 0$. After a few straightforward manipulations we end up with  
 \begin{multline}\label{eq:X-asymptotic}
 \mathcal{X} \doteq -\frac{\phi(0)^{3}}{2\pi\sqrt{\Lambda D}} \left\{\frac{2T}{\sqrt{a}} \int_{0}^{\infty}dx \frac{\sin\left(\arctan(x)/2 \right)}{x\left[1 + x^{2} \right]^{1/4}}
 \right. \\ \left.
+\  \frac{1}{3\pi\Lambda A} \left[\left(a^{2} + 4 \pi^{2}\Lambda^{2}A^{2}t^{2}\right)^{3/4} 
 \right.\right. \\ \left.\left.
-\ \left(a^{2} + 4 \pi^{2}\Lambda^{2}A^{2}T^{2}\right)^{3/4}\right]\right\} \,.
 \end{multline}
The numerical value of the integral in the first term is $\pi/2$. This asymptotic form of the $\mathcal{X}$ integral holds only in the limit $a\to 0$. We compared in Fig.~\ref{fig:y-int-comp} the full, Eq.~\eqref{eq:X-full}, and the asymptotic, Eq.~\eqref{eq:X-asymptotic}, expression of a dimensionless integral $\sqrt{a}\mathcal{X}/\phi(0)^{2}$  as a function of the Kondo parameter $a\in(0,1)$ on the logarithmic scale at a temperature $T=0.1 t$ and with interaction $U=5t$. We can see quite a good agreement in the limit $a\to 0$.

Integral $\mathcal{X}$ contains two contributions. The first one, proportional to $T/\sqrt{a}$, is the dominant critical contribution at nonzero temperatures and prevents the existence of the critical point with $a=0$ signaling the transition to the antiferromagnetically ordered phase. The latter term in the brackets is due to quantum fluctuations dominant at zero temperature where the first term vanishes. It tells us that a transition to a Néel state, $a=0$, is formally possible at zero temperature. But it is nonetheless unphysical, since zero absolute temperature can be reached only asymptotically from finite non-zero temperatures where the long-range order is suppressed by the first term in Eq.~\eqref{eq:X-asymptotic}. 

\subsection{Half-filled band}

We choose the half-filled band with $\bar{\mu} =0$ to determine the explicit temperature dependence  of the Kondo scale $a$. It will directly lead via Eq.~\eqref{eq:Chi-AF-static} to the temperature dependence of the staggered susceptibility.  We have in this symmetric situation
\begin{subequations}
\begin{align}
\Lambda &= - \frac{1 - a}{\phi(0)} \,, \\
A &= \frac {\rho(0)}{8T}  \,, \\
\phi(0) &=  P\int_{-\infty}^{\infty} d\epsilon \rho(\epsilon) \frac{f(\epsilon )}{\epsilon } \nonumber \\
 & \doteq - \rho(0) \ln\left(\frac tT \right)   \,.
\end{align}
\end{subequations}
If we introduce dimensionless parameters $u= -U\phi(0)$ and $y(a)=4\mathcal{X}(a)/\phi(0)^{2}$  the Kondo scale $a$ is then determined from a cubic equation
\begin{equation}\label{eq:a-SC1}
(1 - a)^{3}y(a) + (1 - a) = u\,.
\end{equation} 
The solution of this  equation in the critical region for $a\ll 1$ is $y(a) = u - 1$ with $y\propto T/\sqrt{a}$. The critical region can, however, be reached only in the strong-coupling regime, that is if
\be
U|\phi(0)| \doteq U\rho(0)\ln\left(\frac t T\right)  \gg 1 \,. 
\ee
 There is hence a crossover temperature for arbitrary interaction $U>0$ at which the solution of the half-filled Hubbard model  goes over to a strong-coupling regime. This temperature is 
\begin{equation}\label{eq:T0}
 T_{0} = t \exp\left(- \frac 1{U\rho(0)}\right) \,.
 \end{equation}
The crossover temperature is reminiscent of the Kondo temperature from the $sd$ Kondo model.\cite{Andrei:1983aa}

\begin{figure}
\includegraphics[width=8cm]{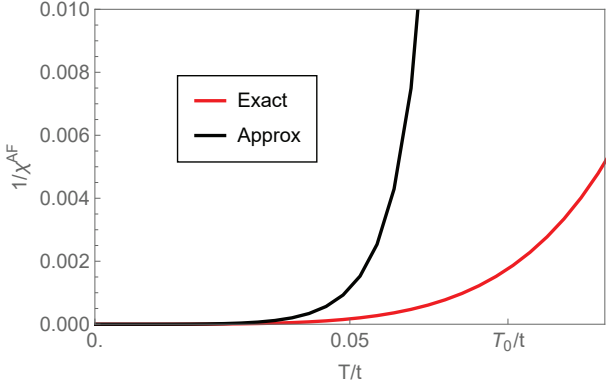}
\caption{Comparison of the temperature  dependence of the staggered susceptibility, Eq.~\eqref{eq:ChiAF-asympt},   calculated for $U=5t$ with the full and asymptotic expressions for the Kondo scale $a$, Eq.~\eqref{eq:Kondo-scale-full}. We indicated the crossover temperature  $T_{0} =0.081 t$ from Eq.~\eqref{eq:T0}.  \label{fig:ChiAF-inv} }
\end{figure}

Finally, the temperature dependence of the Kondo scale in the strong-coupling regime  for $T< T_{0}$  is
\begin{equation}\label{eq:Kondo-scale-full}
a = \frac{4\rho(0)}{D} \frac{T^{2} \rho(0)^{2}\ln^{3}\left(t/T\right)}{\left(U\rho(0)\ln\left(t/T\right) - 1\right)^{2}} \,.
\end{equation}   
The low-temperature asymptotics of the staggered susceptibility for $T\ll T_{0}$ then is
\begin{align}\label{eq:ChiAF-asympt}
\chi^{AF} &\doteq \frac{ D}{2}\ \frac{U^{2}}{T^{2} } \,.
\end{align}
The zero-temperature limit of the half-filled band is a quantum critical point with the divergent staggered susceptibility. We plotted the temperature dependence of its inverse for an intermediate coupling  $U=5t$ in Fig.~\ref{fig:ChiAF-inv} It shows that both its asymptotic limit from Eq.~\eqref{eq:ChiAF-asympt} and the full susceptibility, Eq.~\eqref{eq:Chi-AF-static}, coincide in the critical region sufficiently below the crossover temperature $T_{0}$ from Eq.~\eqref{eq:T0} and approach zero.

\section{Spectral properties}

The one-dimensional Hubbard model at half filling in the limit of zero temperature is in the strong-coupling regime. It means that we cannot apply the standard weak-coupling perturbation theory and the spectral properties at zero temperature must be determined non-perturbatively. The fundamental quantity determining the spectral properties of the correlated electron systems is the spectral self-energy $\Sigma^{Sp}(\veck,\omega_{+})$ renormalizing the dispersion relation and the one-electron propagator from Eq.~\eqref{eq:G-full}. We used the Hartree propagator to determine the thermodynamic two-particle properties of the model, which was necessary to comply with the Ward identity and to keep the approximation with the static effective interaction thermodynamically consistent.\cite{Janis:2019aa} The spectral self-energy, a dynamical correction to the Hartree term,  is generally determined  from the Schwinger-Dyson equation. The only restriction on the spectral self-energy is thermodynamic consistency with the two-particle functions. It means that it should share the critical behavior  determined  from the two-particle vertex to maintain thermodynamic consistency. The Schwinger-Dyson equation must act in the spin-symmetric subspace and use the spin-symmetric two-particle vertex determined in the preceding sections to achieve the demanded thermodynamic self-consistency.     

The Schwinger-Dyson equation with the effective interaction $\Lambda$ as the two-particle irreducible vertex reads  
 \begin{multline}\label{eq:Sigma-Sp}
 \Sigma^{Sp}(k,\omega_{+}) =  - U\Lambda\int_{-\infty}^{\infty}\frac{l_{0}dq}{2\pi}\ P \int_{-\infty}^{\infty} \frac{dx}{\pi}
  \\ 
\times\left\{b(x) \mathcal{G}(k + q, \omega_{+} + x) \Im\left[\frac{\phi(q,x_{+})}{1 + \Lambda \phi(q,x_{+})} \right]     
 \right. \\ \left.
 -\  f(x + \omega)\frac{\phi(q, x_{-})}{1 + \Lambda\phi(q, x_{-}) }\Im \mathcal{G}(q + k,x + \omega_{+})\right\} \,.
 \end{multline}
The bubble $\phi(q,\omega_{+})$ is determined from the thermodynamic propagator $G(k,\omega_{+})$ not to affect the critical behavior of the model with $a\to 0$. The one-electron Green function $\mathcal{G}(k,\omega_{+})$ in this equation is, however the full Green function renormalized via the resulting self-energy as defined in Eq.~\eqref{eq:G-full}. It means that we use the full one-particle renormalization for the one-particle Green function of the Schwinger-Dyson equation determining the spin-symmetric self-energy.  

Equation~\eqref{eq:Sigma-Sp} is solved iteratively starting from the bare (Hartree)  one-electron propagator $G(k,\omega_{+})$. Similarly as in the case of the staggered susceptibility we can perform the integral over momentum $q$ using the expansion around the edge momentum $Q=\pi/l_{0}$ capturing the critical antiferromagnetic fluctuations for $a\to 0$ in the low-temperature limit of the half-filled band. The first iteration for the spectral self-energy then is   
\begin{multline}\label{eq:ImSigma0}
\Im \Sigma^{Sp}_{0}(\epsilon, \omega_{+}) =  \frac{U\phi(0)}{2\pi\sqrt{D}} \left[ b(\epsilon - \omega) + f(\epsilon) \right]
\\
\times  \frac{\sin\left(\alpha(\epsilon - \omega )/2\right)}{\left[ \left(\phi(0) - (1 - a)\Re\phi(\epsilon - \omega)\right)^{2} + (1 - a)^{2}\Im\phi(\epsilon - \omega)^{2}\right]^{1/4}} \,,
\end{multline}   
where the bubble is taken  from Eq.~\eqref{eq:phi-AF} with $\bar{\mu}=0$, the Kondo scale $a$ from Eq.~\eqref{eq:Kondo-scale-full}, and the angle
\begin{align}
\alpha(x) &= \arctan\left(-\frac{\pi\rho(0)^{2}x}{\phi(0)a} \right)\,.
\end{align}
We also used the solution for the effective interaction $\Lambda= (a - 1)/\phi(0)$. The self-energy depends in this mean-field approximation on momentum only via the dispersion relation $\epsilon(k)$ at the critical point. Such a simplified dependence is generally expected for low frequencies around the Fermi energy.  

Already this non-self-consistent solution for the  self-energy qualitatively reveals the expected features of the spectral function at zero temperature. Since the static bubble at half filling is logarithmically divergent at zero temperature, $\phi(0) \propto \rho(0)\ln(t/T)$, the imaginary part of the self-energy can either be zero or infinity. It appears that $\Im \Sigma^{Sp}_{0}(\epsilon, \omega_{+})=\infty$ for $|\epsilon|\le |\omega| \le 2t$. Consequently, $\Im\mathcal{G}(\epsilon,\omega) =0$ and the spectral function has a gap on this interval.  The imaginary part of the self-energy at the edge point $\epsilon =\omega$ has the low-temperature critical asymptotics
\begin{equation}
\Im \Sigma^{Sp}_{0}(\omega, \omega_{+}) \doteq -\ \frac{U^{4}D}{16 T^{2}\sqrt{\ln(t/T)}} \,,
\end{equation}
while the asymptotics for $|\epsilon|<|\omega|$ is slower
\begin{equation}
\Im \Sigma^{Sp}_{0}(\epsilon, \omega_{+}) \doteq -\ \frac{U^{2}\sqrt{\ln(t/T)}}{2\sqrt{2} \pi T} \,.
\end{equation}
The actual critical asymptotics will be modified in the full self-consistent solution of Eq.~\eqref{eq:Sigma-Sp}. The non-self-consistent solution is, nevertheless, sufficient to illustrate the mechanism behind the opening of the gap in the spectral function in this approximation. 

\section{Conclusions}

The ground state and thermodynamics of the one-dimensional Hubbard model can be determined within rigorous approaches. It is, however, very laborious to get an overall picture of the behavior of this model from them. Moreover, the Bethe Ansatz, on which the rigorous results are based, is not extensible to higher dimensions. That is why approximate schemes applicable not only to higher dimensional models but also to more realistic descriptions of interacting electrons are being developed and tested.

We used a universal mean-field-like approximation with a renormalized static effective interaction to study the critical antiferromagnetic fluctuations in the half-filled band. The equilibrium state of our approximation remains paramagnetic down to zero temperature as expected. The low-temperature limit is, however, nontrivial with a crossover temperature below which the system gets into the strong-coupling regime for arbitrary non-zero interaction $U>0$. The solution at zero temperature then is at a quantum critical point with infinite antiferromagnetic fluctuations  and a diverging staggered susceptibility the critical asymptotics of which was derived.  We also showed how a gap in the spectral function opens at zero temperature without being antiferromagnetically  ordered. 

The exact spectrum of the Hubbard Hamiltonian in one dimension reveals that the limit to the non-interacting case $U=0$ is not analytic. One hence cannot rely on the picture of weakly interacting quasiparticles of the Fermi liquid at low temperatures.  Low temperatures can be described within many-body approaches only via nonperturbative approximations renormalizing sufficiently the bare interaction enabling a smooth extension to strong coupling. Our main conclusion is that the zero-temperature state is strongly correlated and can be approached within the standard many-body techniques with fermionic excitations only asymptotically from high temperatures. Approximate solutions derived within the many-body perturbation theory at strictly zero temperature cannot be trusted in low-dimensional systems ($d<3$).    
       
\section*{Acknowledgment}
 The research was supported by Grant No. 19-13525S of the Czech Science Foundation and INTER-COST LTC19045 of the Czech Ministry of Education, Youth and Sports. 
 
 \section*{Data Availability}
 
We have not used any specific data needed to reproduce  the conclusions of this study beyond that available in the article.


\end{document}